\documentclass[trackchanges,twocolumn,twocolappendix]{aastex7}

\usepackage{amsmath}
\usepackage{graphicx}
\usepackage{comment}

\begin{document}

\title{Synchronous Rotation in the (120347) Salacia-Actaea System}

\author[0009-0004-7149-5212]{Cameron Collyer}
\affiliation{Florida Space Institute, University of Central Florida, 12354 Research Parkway, Orlando, FL 32826, USA}
\affiliation{Department of Physics, University of Central Florida, 4111 Libra Drive, Orlando, FL 32816, USA}
\email[show]{cpcollyerastro@gmail.com}  

\author[0009-0004-7149-5212]{Estela Fernández-Valenzuela}
\affiliation{Florida Space Institute, University of Central Florida, 12354 Research Parkway, Orlando, FL 32826, USA}
\email{}

\author[0000-0002-8690-2413]{Jose Luis Ortiz}
\affiliation{Instituto de Astrofísica de Andalucía, Glorieta de la Astronomía s/n, 18008 Granada, Spain}
\email{}

\author[0000-0002-6117-0164]{Bryan J. Holler}
\affiliation{Space Telescope Science Institute, Baltimore, MD 21218, USA}
\email{bholler@stsci.edu}

\author[orcid=0000-0002-1788-870X,sname='Proudfoot']{Benjamin Proudfoot}
\affiliation{Florida Space Institute, University of Central Florida, 12354 Research Parkway, Orlando, FL 32826, USA}
\email{benp175@gmail.com}

\author[0000-0003-0419-1599]{Nicolás Morales}
\affiliation{Instituto de Astrofísica de Andalucía, Glorieta de la Astronomía s/n, 18008 Granada, Spain}
\email{}

\author[0000-0003-1661-0594]{Rafael Morales}
\affiliation{Instituto de Astrofísica de Andalucía, Glorieta de la Astronomía s/n, 18008 Granada, Spain}
\email{}

\author[0000-0001-8821-5927]{Susan Benecchi}
\affiliation{Planetary Science Institute, 1700 East Fort Lowell, Suite 106, Tucson, AZ 85719, USA}
\email{}

\author[0000-0002-6085-3182]{Flavia L. Rommel}
\affiliation{Florida Space Institute, University of Central Florida, 12354 Research Parkway, Orlando, FL 32826, USA}
\email{}

\author[orcid=0000-0002-8296-6540, sname='Grundy']{Will Grundy} 
\affiliation{Lowell Observatory, 1400 W Mars Hill Rd, Flagstaff, AZ 86001, USA}
\affiliation{Northern Arizona University, Department of Astronomy \& Planetary Science, PO Box 6010, Flagstaff, AZ 86011, USA}
\email{grundy@lowell.edu}

\author[orcid=0000-0003-1080-9770, sname='Ragozzine']{Darin Ragozzine} 
\affiliation{Brigham Young University Department of Physics \& Astronomy, N283 ESC, Brigham Young University, Provo, UT 84602, USA}
\email{darin_ragozzine@byu.edu}

\begin{abstract}
\noindent We report on roughly 16 years of photometric monitoring of the transneptunian binary system (120347) Salacia-Actaea which provides significant evidence that Salacia and Actaea are tidally locked to the mutual orbital period in a fully synchronous configuration. The orbit of Actaea is updated, followed by a Lomb-Scargle periodogram analysis of the ground-based photometry which reveals a synodic period similar to the orbital period and a peak-to-peak lightcurve amplitude of $\Delta m$ = 0.0900 $\pm$ 0.0036 mag (1$\sigma$ uncertainty). Incorporating archival HST photometry that resolves each component, we argue that the periodicity in the unresolved data is driven by a longitudinally varying surface morphology on Salacia, and derive a sidereal rotation period that is within 1$\sigma$ of the mutual orbital period. A rudimentary tidal evolution model is invoked that suggests synchronization occurred within 1.1 Gyr after Actaea was captured/formed. 
\end{abstract}

\section{Introduction}
\label{introduction}
Much of what we know about the interiors of transneptunian objects (TNOs) is derived from detailed study of binary systems. Constraining the mutual orbit of a binary system (yielding a total system mass) and estimating the volume by means of stellar occultations or thermal radiometry gives a first order estimate of its bulk composition and bulk mass-density (density hereafter). Observations of TNOs point to a size-density relationship where density increases with size \citep[e.g.,][]{Brown2012, Grundy2019UK}. This has been interpreted as being the result of the compaction of a rock-rich core and its drop in porosity during thermal evolution and differentiation, an effect that increases with increasing size, mass and rock mass fraction \citep{Bierson2019, Loveless2022}. The spin-orbit configurations of the largest TNO-satellite systems have been reproduced by simulations of giant impacts, where the satellite is captured as an intact fragment of the impactor or forms from a collisional debris disk around the primary \citep[e.g.][]{Arakawa2019}.

Here we focus on the TNO binary system (120347) Salacia-Actaea. Salacia-Actaea is dynamically classified as a hot classical, main belt TNO \citep{Gladman2021}. The surface compositions of Salacia and Actaea are likely quite similar. Resolved HST multi-band photometry show they have similar optical colors \citep{Benecchi2009}. The color-albedo dichotomy of TNOs \citep{Lacerda2014} also suggests the low geometric albedo derived for the unresolved system from radiometry \citep{Vilenius2012,Fornasier2013,Brown2017} is representative of both surfaces. The shallow spectral slope of each components' surface and the low geometric albedo is characteristic of the BrightIR TNO surface class \citep[e.g.,][]{Fraser2surfaces}. The dynamically excited orbit and BrightIR surface class point to the system forming closer to the sun and being scattered outwards onto its current orbit during Neptune's migration. The diameters of Salacia and Actaea have been measured through radiometry and the relative difference in their optical reflectance to be 866 $\pm$ 37 km and 290 $\pm$ 21 km, respectively \citep{Brown2017}. Using the system mass (which we update in this work, see Table \ref{tab:fits}) and total system volume, the effective density of the system is $\rho = 1.38^{+0.22}_{-0.18}$ g cm$^{-3}$. This density appears to be intermediate between low mass, highly porous worlds and the dwarf-planet class where significant central-compaction is expected. Other TNO worlds in this intermediate size-density regime are (90482) Orcus, (38628) Huya, (174567) Varda, (55637) 2002 UX25, and (229762) G!k{\'u}n$||$'h{\`o}md{\'i}m{\`a}.

In this work, we present observational evidence that supports the Salacia-Actaea system being fully tidally evolved today, where lightcurve analysis points to Salacia rotating at the mutual orbital period. We then invoke the tidal evolution model of \cite{Goldreich1966}, which predicts total synchronization of Salacia-Actaea within the age of the solar system if Salacia's interior is within an order of magnitude as dissipative as the interior of (136199) Eris. In light of these results we propose other TNO systems that may also be doubly synchronous, and suggest future work that could further constrain the properties of TNO interiors in the intermediate-density regime. 

\section{The Mutual Orbit}
\label{orbit0}
In this section we update the orbit for the Salacia-Actaea system which was last updated based on observations in 2016 \citep{Grundy2019}. To do this, we acquired 3 new resolved observations of Salacia and Actaea, one with Keck in 2021 and two with HST in 2024, increasing the observational baseline by 8 years. Our full astrometry dataset is shown in Table \ref{tab:observations}. Orbit fitting was completed using MultiMoon, an orbit fitter built for TNO binary orbit fitting \citep{ragozzine2024beyond,proudfoot2024beyond}. The dataset we used was taken verbatim from \citet{Grundy2019}, although with our added observations. See Appendix \ref{orbit1} for further details regarding these observations and the orbit-fitting procedure. Our derived orbit is nearly identical to the most recent found in the literature \citep{Grundy2019}, although the recovered parameters are somewhat more precise.

Two results of particular relevance here are the updated system mass and mutual orbital period, being $M_{sys} = 486.1^{+7.6}_{-7.4} \times 10^{18}$ kg and $P_{orb} = 5.49389 \pm 0.00001$ days, respectively (See Table \ref{tab:fits} for full Keplerian orbit solution). Using the diameters of Salacia and Actaea derived in \cite{Brown2017} (866 $\pm$ 37 km and 290 $\pm$ 21 km, respectively) to calculate the total system volume, this yields an effective bulk density of $\rho = 1.38^{+0.22}_{-0.18}$ g cm$^{-3}$ for the system. This density is roughly 10\% larger than the bulk density derived in \cite{Brown2017} due to the larger updated system mass.

\section{Lightcurve Observations and Photometry}
\subsection{Description of Observations}
\label{observations}

The ground-based photometry used in this work were obtained from August 2005 - October 2021. The total on-target time was $\sim$165 hr, or 6.90 days. Roughly 20\% of observations were made before 2010, 60\% between 2011 - 2015 and 20\% after January 2020. Observations were made with 11 different telescope/camera/filter combinations. A description of the hardware used along with the number of observing runs and images for each telescope/camera/filter combination is tabulated in Table \ref{obstable}.

\begin{deluxetable}{llllllllllll}
\tablecaption{Observation Description \label{obstable}} 
\tablehead{Telescope & Diameter & CCD Camera & Bin & $x$ $\times$ $y$ & $s$ & FOV & Filter & $N_O$$^{(a)}$ & $N_m$$^{(b)}$ & Date \\
Name & (m) & & &  (px $\times$ px) & ($''/$px) &  & & & & Span }

\startdata
CAHA$^{(c)}$  & 1.2 & DLR-MKIII  & 2$\times$2  & 2048$\times$2056  & 0.63 & 21.4’$\times$21.4’ & $O$   & 8  & 148   & 2014.96 - 2021.78   \\
CAHA$^{(c)}$  & 2.2 & SITe-1d    & 2$\times$2  & 1024$\times$1024  & 0.53 & 9’$\times$9’       & $R$   & 6  & 10    & 2011.73 - 2012.79   \\
              &     &            &             &                   &      &                    & $V$   & 6  & 8     & 2011.73 - 2012.79   \\
              &     &            &             &                   &      &                    & $KG$  & 7  & 27    & 2011.66 - 2011.83   \\
CAHA$^{(c)}$  & 3.5 &  SITe-16a  & 2$\times$2  & 1000$\times$1000  & 0.64 & 11’$\times$11’     & $R$   & 1  & 1     & 2014.55 - 2014.55   \\
              &      &           &             &                   &      &                    & $V$   & 1  & 2     & 2014.55 - 2014.55   \\
NOT$^{(d)}$   &  2.6 & ALFOSC    & 1$\times$1  & 2048$\times$2064  & 0.21 & 6.4’$\times$6.4’   & $R_B$ & 1  & 4     & 2011.66 - 2011.66   \\
OSN$^{(e)}$   &  1.5 & Roper T150& 2$\times$2  & 1024$\times$1024  & 0.46 & 7.9’$\times$7.9’   & $O$   & 45 & 572   & 2005.59 - 2021.76   \\
TNG$^{(f)}$   &  3.6 & DOLORES   & 2$\times$2  & 1050$\times$1050  & 0.51 & 8.6’$\times$8.6’   & $I$   & 2  & 8     & 2011.50 - 2011.51   \\
              &      &           &             &                   &      &                    & $R$   & 10 & 194   & 2011.50 - 2011.83   \\
              &      &           &             &                   &      &                    & $V$   & 4  & 14    & 2011.50 - 2011.51   \\
\enddata
\tablecomments{$(a)$ Number of individual observing runs. $(b)$ Cumulative number of photometric datapoints from all observing runs. $(c)$ Calar Alto Astronomical Observatory, Spain. $(d)$ Nordic Optical Telescope, Canary Islands. $(e)$ Sierra Nevada Observatory, Spain. $(f)$ Galileo National Telescope, Canary Islands.}
\end{deluxetable}

\subsection{Absolute Magnitude Calibration}
\label{photometry0}

\begin{figure}
\centering
 	\includegraphics[scale=0.62]{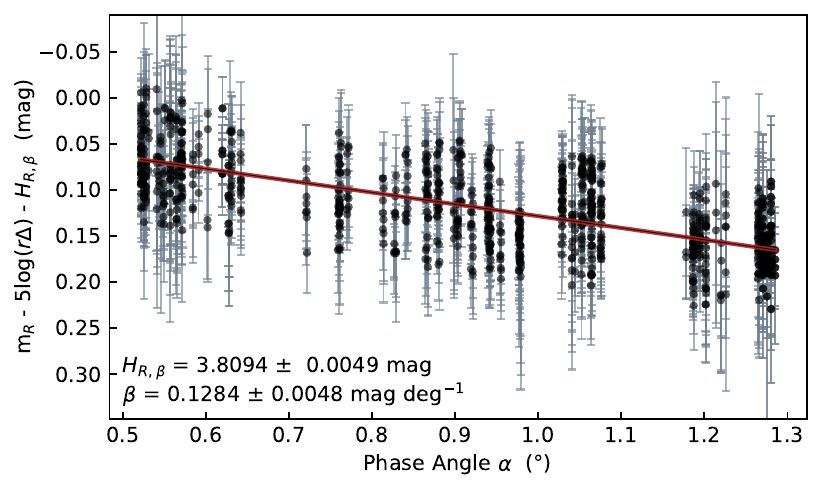}
	\caption{
The magnitude dependence as a function of the Sun-TNO-Earth phase angle $\alpha$. Over the $0.5^{\circ} \lesssim \alpha \lesssim 1.3^{\circ}$ range the relationship is clearly linear.
 \label{figure:phase curve}
 }
\end{figure}

We utilized the Massive prOcessing Of aStronomical imagEs (Moose) - version 2 \citep[M2;][]{M2} software to process FITS images from each observing run and output photometry of the Salacia-Actaea system. Photometry is calibrated to the Johnson-Cousins $R$-band on the \cite{Landolt1992} photometric system. In Appendix \ref{photometry1} we describe the M2 software and the method used to transform the photometry.

The heliocentric and geocentric distances and Sun-TNO-Earth phase angle\footnote{Heliocentric and geocentric distances and Sun-Earth-TNO phase angles were queried from the JPL SBDB (\url{https://ssd.jpl.nasa.gov/tools/sbdb_lookup.html\#/?sstr=salacia}) using Astroquery \citep{astroquery2019}.} varied over the course of the $\sim$16 years of observations. This would obscure any photometric periodicity we look to measure from the system, and so we needed to transform to a constant geometry. The absolute magnitude $H$ of a small body is the magnitude it would have if it were located at the (impossible) geometry $\Delta = r = 1$ au and a phase angle of $\alpha = 0^{\circ}$, where $\Delta$ is the geocentric distance and $r$ is the heliocentric distance. In order to remove the phase-angle dependence on the photometry, we constrained the phase curve of the Salacia-Actaea system for phase angles $\alpha$ in the range $0.5^{\circ} \lesssim \alpha \lesssim 1.3^{\circ}$. Over this range of phase angles, we noted the phase curve of the Salacia-Actaea system is virtually consistent with a linear trend, so we chose a simple linear model for the phase curve. When the phase function contribution is approximated as linear, the $H$-magnitude equation is 

\begin{equation}
    H = m - 5\text{log}(r\Delta) - \beta \alpha,
    \label{eq: Hbeta}
\end{equation}

\noindent where $\beta$ is the linear phase coefficient with units mag deg$^{-1}$ and $m$ is the apparent magnitude ($R$-band). We highlight that the $H_R$ reported here for Salacia-Actaea is calculated without constraints on the phase-function at the lowest phase-angles where we may be missing a non-linear opposition surge \cite[e.g.,][]{Verbiscer2022}. This $H_R$ magnitude may therefore be an over-estimate of the true $H_R$, and so we refer to it as $H_{R,\beta}$. Though this may be true, it has no effect on our period search since it removes all relative phase-angle dependence. We find $H_{R,\beta}$ and $\beta$ to be 3.8094 $\pm$ 0.0049 mag and 0.1284 $\pm$ 0.0048 mag deg$^{-1}$, respectively, as shown in Figure \ref{figure:phase curve}\footnote{The phase curve, $H_{R,\beta}$ and $\beta$ fit parameters explicitly presented here were evaluated after the analysis in the following section. The photometric periodicity was removed. See Appendix \ref{photometry1} that explains this iterative procedure.}.

If the photometric phase is parametrized by the variable $\theta$, where $\theta \in [0,1]$, we define the $H$-magnitude as the period-average of 

\begin{equation}
    h(\theta) = H + f(\theta),
\end{equation}

\noindent where $f(\theta)$ represents the period-phase dependence with a period-average of 0. In analyzing the absolute magnitude of the system over $\sim$16 yr of the observations we find no evidence of a systematic change in $H_{R,\beta}$.



\section{Period Search}
\label{periodsearch}
We conducted a period search on the photometric data ($h$-mags) using Astropy's \texttt{Time Series LombScargle} class \citep{astropy:2013, astropy:2018, astropy:2022}, which invokes the Lomb-Scargle \citep{Lomb1976, Scargle1982} periodogram (periodogram hereafter) on time series data and takes uncertainties into account. We searched for sinusoidal signals in the periodogram analysis; a peak in the periodogram at frequency $f_i$ suggests that there is a photometric signal from the data of the form $h_R = H_R + A\text{cos}(2 \pi f_i t + \phi)$. The following periodogram analysis follows the guiding work of \cite{VanderPlas2018}. 

\subsection{Periodogram of the Window Function}
\label{windowing}

We first analyzed a periodogram of the time-sampling (window function) to constrain how the observation windowing distorts the periodogram of the photometry. Any unique frequencies with nonzero powers in the window-periodogram (1) may show up as false peaks in the periodogram of the photometry and (2) creates aliases of true photometric frequencies exhibited by the Salacia-Actaea system. If Salacia-Actaea exhibits a real photometric frequency at $f_{\text{real}}$, then each window frequency $f_{\text{w}}$ produces aliases of $f_{\text{real}}$ along the sequences

\begin{equation} \label{eq:alias}
\begin{split}
 f_{\text{real}} \pm kf_{\text{w}}, \hspace{3.4cm} \\
-f_{\text{real}} \pm kf_{\text{w}}, \hspace{0.5cm} k = 1,2,3,...,\pm\infty.
\end{split}
\end{equation}

\noindent When assessing the periodogram of the photometry itself, peaks in the power spectrum corresponding to frequencies along these sequences are therefore derived from the true driving frequency $f_{\text{real}}$ and are not unique \citep{VanderPlas2018}.

We created the window function by setting all values of the photometry to unity. In evaluating the window periodogram we did not pre-center the data or use a floating mean. The periodogram of the window function revealed that the main frequencies arising from the time-sampling were (1) Earth's sidereal rotation period ($\sim$0.9972 days) and (2) the synodic period of the moon ($\sim$29.527 days). 

\subsection{Periodogram of the Salacia-Actaea Photometry}
\label{photometryperiodogram}
\begin{figure*}
\centering
        \includegraphics[width = \textwidth]{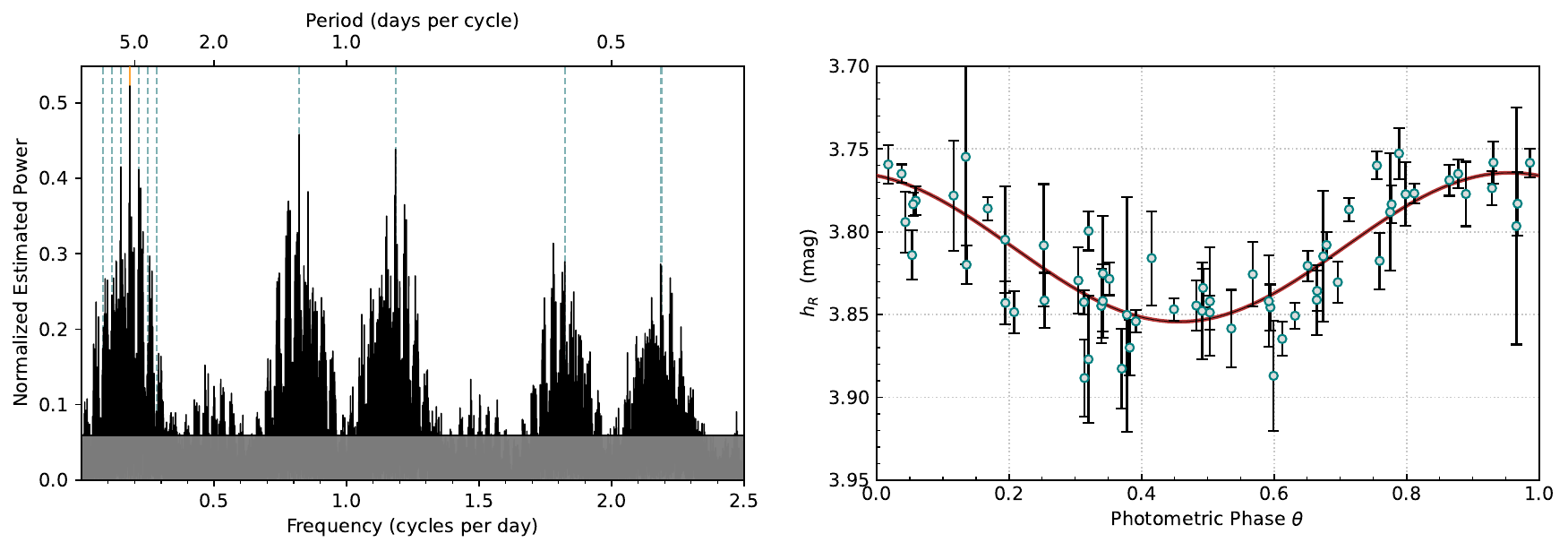}
	\caption{Left: Lomb-Scargle periodogram of the unresolved ground-based photometry of the Salacia-Actaea system in the range where spectral powers are significant. The frequency with the highest power, $\sim$0.182 cycles-per-day, is highlighted in orange. The blue dashed lines are the expected locations of aliases of this frequency (up to $k=3$) due to the windowing frequencies (See Section \ref{windowing}). Right: The time-series photometry ($h$-mags vs. time) folded over the synodic period $T_{\text{syn}}$ = 5.49430 days. Since this period is much larger than the time-span of any single observing run when photometry was measured, we plot the weighted average of each observing run and its uncertainty. The fact that multiple observing runs which sample the same rotational phase are in agreement bolsters confidence that this period is legitimate. The solid red line is the best-fit phase-shifted sinusoid.
 \label{figure:periodogramphasefold}
 }
\end{figure*}
We now present the results of the periodogram analysis of the unresolved ground-based photometry of the Salacia-Actaea system. An initial power spectrum using a frequency grid with an upper bound of 1 cycle-per-hour showed that frequencies above 2.5 cycles-per-day corresponded to very small normalized powers, and so we focused on frequencies lower than 2.5 cycles-per-day, corresponding to periods greater than 9.6 hour. The 6.5 hour photometric period of the system published in \cite{Thirouin2014} did not appear to be a driving frequency in our periodogram analysis, though it is close to the $k=4$ alias of the main frequency found below (see also Section \ref{compareThirouin}).


In order to evaluate the noise floor of the spectral power in the periodograms, we bootstrap re-sampled the photometry by randomly re-assigning $h$-mags and their uncertainties to the timing array. We did this 10000 times. For each bootstrap run, the maximum power achieved was recorded. In Figure \ref{figure:periodogramphasefold} (left), the power corresponding to the height of the grey box is the largest normalized estimated power achieved in the bootstrapping runs, where peaks above this noise floor have a 0.1\% false-positive rate \citep{VanderPlas2018}. Spectral powers above this line, generally speaking, are therefore caused by true photometric periodicities and their aliases with 99.9\% confidence. 

The frequency corresponding to the highest power in the periodogram was $\sim$0.182 cycles-per-day, with an equivalent period of $\sim$5.494 days. This is similar to Actaea's orbital period of 5.49389 $\pm$ 0.00001 days (Table \ref{tab:fits}). We performed 10000 Monte-Carlo runs of the periodogram by re-sampling each photometric data point from a gaussian centered at the measured $h$-mag and scaled by its uncertainty. In all 10000 runs, the same frequency of 0.182 cycles-per-day showed the highest spectral power. The main structures seen in this periodogram correspond to aliases of the 0.182 cycles-per-day frequency, and aliases of those aliases. 

\subsection{Phase Folded Lightcurve}
In Figure \ref{figure:periodogramphasefold} (right) we folded the time-series photometry by the $\sim$5.494 day period, which revealed an underlying signal with a sinusoidal shape. Since this period is much larger than the time-span of any single observing run when photometry was measured, we plot the weighted average of each observing run and its uncertainty. Note that the periodogram analysis of the previous section was done on the non-binned photometry.

The best-fit synodic photometric period $T_{\text{syn}}$ and peak-to-peak amplitude $\Delta m$ with their corresponding $1\sigma$ uncertainties are $T_{\text{syn}}$ = 5.49430 $\pm$ 0.00016 days and $\Delta m$ = 0.0900 $\pm$ 0.0036 mag. Gaussian uncertainties for the period as well as for the amplitude of the lightcurve were determined as follows, using the period as an example. We used a hyper-fine grid of trial periods (finer than the periodogram grid-spacing) around the best-fit period and stepped through each period, phase folding the data by this period and fitting a phase-shifted cosine function, allowing the amplitude to be a free parameter. The uncertainty in the period corresponds to the difference between the best-fit period and the period where the minimum $\chi^2$ increased by unity. The same was done for an amplitude grid centered on the best-fit amplitude. We find the uncertainties on each parameter to be symmetric. In the following sections we discuss the nature of this periodicity.

\subsection{Considering Previous Results for Rotation}
\label{compareThirouin}
As noted above, the photometric periodicity of $T_{\text{syn}}$ = 5.49430 $\pm$ 0.00016 days is discrepant with the results of \cite{Thirouin2014} (T14 hereafter) who found $T_{\text{syn}} \simeq$ 6.5 hours. We would like to highlight that the ground-based photometry dataset used in the present work is 4x as large as the dataset published in T14. Moreover, we computed absolute photometry on the images used in T14 which were incorporated into our dataset (relative photometry was used in T14). The relative photometry used in T14 to search for periodicity in the Salacia-Actaea system is biased towards detecting short-period signals less than the duration of their observing runs (1--4 days). In comparison, the absolute photometry and larger dataset used in the present work is more sensitive to longer-period fluctuations and allows multi-year observations to be combined with greater accuracy in the zeropoint of the photometry.

We also downloaded the photometry from T14 and ran a periodogram analysis. To reproduce the results of T14, we needed to remove the uncertainties when generating the periodogram; when uncertainties were included we found that the 5.49430-day period had the highest estimated power in the spectrum. The relative photometry was phase folded by periods of 5.49430 days and 6.5 hours and we calculated the $\chi^2$ per degree of freedom for each case, yielding values of 1.17 and 1.42, respectively. The smaller reduced $\chi^2$ for the 5.49430-day period is consistent with the results from the periodogram utilizing the uncertainties, as well as the results from Section \ref{photometryperiodogram}. The 6.5-hour period is also close to the $k=4$ alias of 5.49430 days. We conclude that the 6.5 hour period found in T14 is the result of their photometry technique being biased towards detecting short-period signals, and that they likely detected a short-period alias of the true 5.49430-day synodic period.

\section{Synchronous Rotation}

\subsection{Resolved HST Observations}
\label{hstdata}
\begin{figure}
\centering
 	\includegraphics[width=0.48\textwidth]{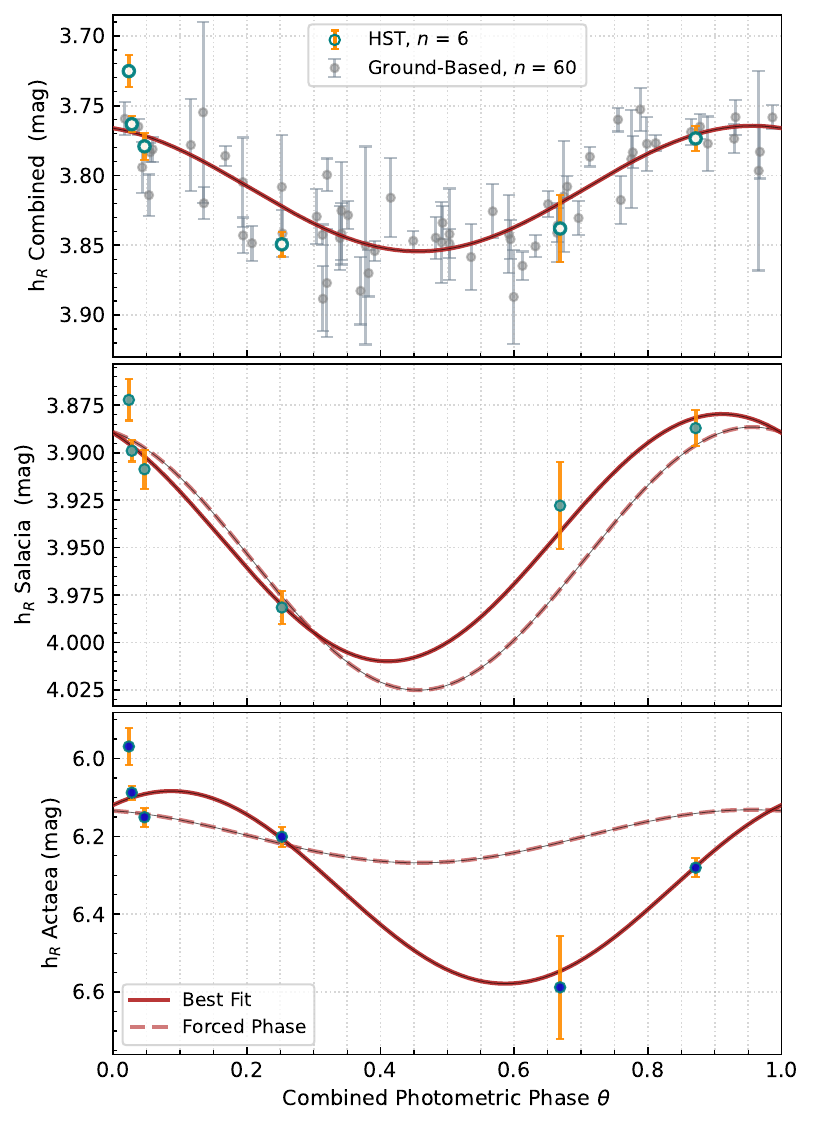}
	\caption{Top: In grey are the ground-based data from Figure \ref{figure:periodogramphasefold}. Data in orange/blue are the combined HST data of the system which clearly match the ground-based data. The solid line model is the best-fit phase-shifted sinusoid. Middle: Resolved HST photometry of Salacia. The dashed line is a sinusoid fit to the data where the phase-offset was set to match that of the unresolved data in the top panel. The solid line is another phase-shifted sinusoid with all free parameters. Bottom: Resolved HST photometry of Actaea. The style of the lines match the model-types applied to the Salacia data.
 \label{figure:hubbledata}
 }
\end{figure}

In order to understand the nature of the photometric signal that appears to oscillate with the same period of Actaea's orbit, we analyzed archival HST data of the system that resolved Salacia and Actaea individually. We analyzed HST images obtained with the Planetary Camera (PC) of the Wide Field and Planetary Camera 2 (WFPC2; \cite{McMaster2008}) that were part of the HST Cycle 16 program 11178, P.I. Will Grundy\footnote{The HST data presented in this article were obtained from the Mikulski Archive for Space Telescopes (MAST) at the Space Telescope Science Institute. The specific observations analyzed can be accessed via \dataset[doi: 10.17909/0ky9-sp71]{https://doi.org/10.17909/0ky9-sp71}.}. The dates of these images span 2007 July 15 - 2008 May 19. Resolved PSF photometry was performed on data taken in the F606W filter following the method described in \cite{Benecchi2009}. Collectively there are $n=6$ photometric points that sample Salacia and Actaea's individual lightcurves. Each point represents the weighted average and the standard deviation on the mean of photometry done on individual frames; there were 4 dither positions for each visit, although one point is comprised of 5 individual frames taken over two consecutive HST orbits for which F606W$-$F814W color data was measured. F606W magnitudes were converted to $R$-band magnitudes by combining the resolved HST photometry and fitting a F606W$-R$ color offset between the HST data (F606W-band) and the model fitted to the unresolved data ($R$-band) (Figures \ref{figure:periodogramphasefold} \& \ref{figure:hubbledata}). We assume this color for Salacia and Actaea, which is a safe assumption given their almost identical F606W$-$F814W colors \citep{Benecchi2009}. Their apparent magnitudes were converted to $h_R$ magnitudes using the phase curve from Section \ref{photometry0}.


In Figure \ref{figure:hubbledata}, the resolved HST data converted to $h_R$ magnitudes is presented in the lower two panels. The epoch of these HST observations was within the time-span of the ground-based observations, and so we folded the time-series data by $T_{\text{syn}}$. The period-phase uncertainty for each datapoint is smaller than the width of the symbols. In the top panel, we combined the resolved HST photometry and plotted this over the ground-based photometry, which shows a close agreement. 

At the outset, we highlight that a photometric periodicity consistent with the mutual orbit period observed in the periodogram analysis is unlikely to be caused by either Salacia or Actaea being triaxial ellipsoids. The lightcurve shape of a triaxial ellipsoid over a full rotation period is `double peaked', meaning there are two minima and two maxima per full rotation. Taking photometry of such a triaxial ellipsoid and performing a period search as described in Section \ref{periodsearch}, the periodogram would show a peak at a frequency equivalent to one-half of the full rotation period, as it searches for sinusoidal signals. If we consider that the observed photometric periodicity of the system is due to an elongated shape, then the true rotation period of the body would be twice the mutual orbit period, such that it would be in the 1:2 spin-orbit resonance. As described in Section \ref{tidalevolution}, it is likely that the Salacia-Actaea system has widened over time due to tidal evolution, where Salacia was spun-down while the semimajor axis increased. If Salacia or Actaea is in the 1:2 spin-orbit resonance, it would have had to pass the 1:1 spin-orbit resonance (assuming the initial spin period was faster than $\sim 5.494$ days), which is unlikely. Therefore the lightcurve amplitude is most likely due to a non-uniform albedo on the surface of Salacia and/or Actaea, which we discuss below.

Since Actaea is much smaller and much less massive than Salacia, Actaea's synchronization timescale (the time it takes for Actaea's primordial rotation period to synchronize with the orbital period) is shorter than that of Salacia. It is expected that Actaea's orbit and spin period are synchronized today (see Section \ref{tidalevolution}). We therefore first tested the hypothesis that only Actaea exhibits synchronous rotation and drives the periodicity seen in the ground-based data. Under this hypothesis, Actaea's lightcurve shape is single-peaked and the observed photometric periodicity is due to a nonuniform surface morphology over the longitudinal zones such as a nonuniform albedo. We first assume that Salacia's lightcurve is constant at the weighted average of its resolved HST photometry, and that Actaea's mean flux contribution is well approximated by the weighted average of its resolved HST photometry. Under these assumptions, Actaea's lightcurve amplitude must be $\Delta m \simeq 0.95$ mag for the combined photometry to exhibit a $\Delta m \simeq 0.09$ mag (Figure \ref{figure:periodogramphasefold}). This lightcurve amplitude translates to a large albedo dichotomy across Actaea's surface. Due to the non-equatorial aspect at which we view Actaea, assuming Actaea's spin pole is closely aligned with the orbit pole direction, a substantial portion of Actaea's surface near the pole is always visible. This increases the longitudinal albedo contrast required to match the lightcurve amplitude. If we assume Actaea's minimum albedo is the low system albedo derived from unresolved radiometry, $p_{\text{min}} \simeq 0.04$, $p_{\text{max}} \simeq 0.14$ yields a $\Delta m \simeq 0.95$ mag. While these albedos are not extreme, such a dichotomy across a small-size TNO like Actaea is difficult to reconcile even with a curated collisional story. 


Moreover, if Actaea is responsible for the observed photometric periodicity, then its period-folded HST photometry must oscillate in-phase with the unresolved photometry. We fitted a simple sinusoidal model to the Actaea photometry by forcing it to be in phase with the ground-based photometry. This resulted in a poor fit with a $\Delta m \simeq 0.18$ mag, significantly smaller than the $\Delta m \simeq 0.95$  mag required to match the observations (dashed line in bottom panel of Figure \ref{figure:hubbledata}). Salacia's lightcurve is also clearly not constant as assumed, which in turn would require a $\Delta m > 0.95$ for Actaea's lightcurve to wash out Salacia's oscillating signal. Collectively, these arguments do not support the hypothesis that the photometric periodicity observed in the ground-based data is due to Actaea alone. 

We then turn to the hypothesis that Salacia and Actaea are doubly-synchronous\footnote{Since Actaea's synchronization timescale is much shorter than Salacia's, it follows that if Salacia's rotation is synchronized, Actaea's is as well.}, where Salacia's lightcurve is single peaked and drives the photometric periodicity observed in the ground-based data. We followed the same methodology described above. A simple sinusoidal model fitted to Salacia's HST photometry (solid line in middle panel of Figure \ref{figure:hubbledata}) resulted in a good fit that is nearly in-phase with the ground-based signal and a $\Delta m \simeq 0.1$ mag. Assuming Salacia's surface has a hemispheric albedo dichotomy this yields an albedo difference of $\sim 5$\%; concentrating an `albedo spot' to a smaller area results in a larger contrast. A similar model forcing the phase-offset to be equal to that of the ground-based data resulted in an adequate fit (dashed line). The simplest explanation for the photometric periodicity found in the unresolved data is therefore that Salacia's longitudinal zones are moderately nonuniform in albedo. The true lightcurve shapes of Salacia and Actaea are more complex than the simple sinusoid models presented here. However, use of a more complicated model is unjustified with only $n=6$ datapoints. Nevertheless, the combined lightcurve shape is clearly well approximated by this simple model. These
results support the hypothesis that Salacia and Actaea exhibit doubly-synchronous rotation.

\begin{figure*}
\centering
 	\includegraphics[scale = 0.65]{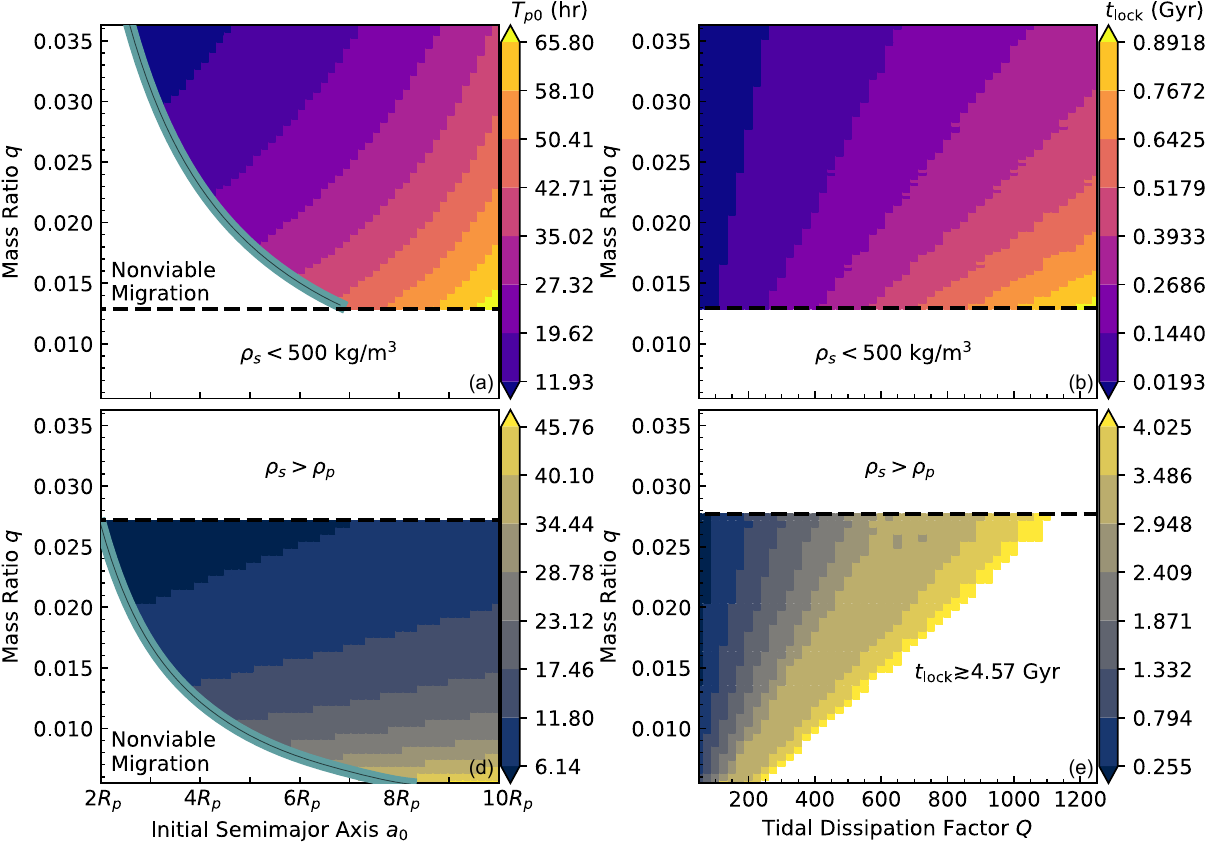}
        \includegraphics[scale = 1]{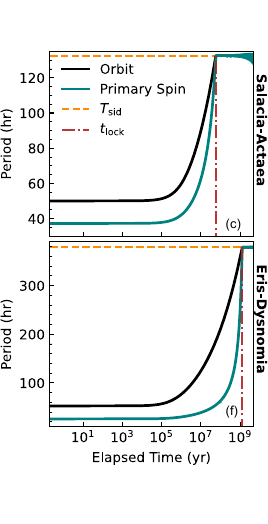}
	\caption{The top row is for the Salacia-Actaea system. (a): $a_0-q-T_{p0}$ space where outward migration is viable by conservation of angular momentum. The range of $q$-values are calculated by allowing Actaea's density to range from 0.5 g cm$^{-3}$ up to the effective system density of $\rho = 1.38^{+0.22}_{-0.18}$ g cm$^{-3}$. (b) Three dimensional plot that shows the dependence of the tidal locking time $t_{\text{lock}}$ on $Q$ and $q$. (c) A schematic example of an integration where Actaea migrates outward while spinning down Salacia. In this example tidal evolution ceases at $t_{\text{lock}} \simeq 50$ Myr. The bottom panels are the same figures but represent the Eris/Dysnomia system. 
 \label{figure:tidalevolution}
 }
\end{figure*}

\subsection{The Sidereal Rotation Period}
The synodic photometric period found in the unresolved ground-based photometry therefore corresponds to the mutual synodic rotation period of Salacia and Actaea ($T_{\text{syn}}$ = 5.49430 $\pm$ 0.00016 days). The orbital period of Actaea, 5.49389 $\pm$ 0.00001 days (Table \ref{tab:fits}), is a sidereal period. The sidereal rotation period $T_{\text{sid}}$ can be estimated from $T^{-1}_{\text{sid}} = T^{-1}_{\text{syn}} +  \frac{\omega_{\text{sys}}}{2 \pi}$, where $\omega_{\text{sys}}$ is the representative heliocentric orbital angular velocity of the system during the observations ($\sim$0.000569 rad/day). We find the sidereal rotation period to be $T_{\text{sid}}$ = 5.49403 $\pm$ 0.00016 days, which is consistent with Actaea's sidereal orbital period within 1$\sigma$. This is only the third TNO-binary system observationally confirmed to exhibit doubly-synchronous rotation; the Pluto-Charon \citep{Christy1978} and Eris-Dysnomia systems \citep{Szakats2023, Bernstein2023} are also in this tidal end-state.


\section{Tidal Evolution}
\label{tidalevolution}
Doubly-synchronous rotation is the end-state of tidal evolution, where angular momentum transfer between rotation and orbit seizes. The present-day rotation period of $\sim 5.49$ days found is very slow compared to typical rotation periods of singleton TNOs ($\sim 10$ hr \citep[e.g.][]{MikeArotate2019, Ashton2023}). It is therefore likely that the orbital separation between the two components was smaller in the past and expanded to its current value ($\sim13.2$ Salacia-radii, $R_p$) through the transfer of angular momentum from Salacia's spin to the orbit. In Appendix \ref{tidalmethods} we present equations that govern the evolution of Salacia's spin period and the orbital separation of the system (the orbital period). The purpose of this section is to explore the range of initial conditions (primaricentric semi-major axes $a_0$ and Salacia spin periods $T_{p0}$), mass ratios $q$, and tidal parameters that allow the system to synchronize within the age of the solar system. In the following we assume Actaea's spin is always synchronized to the orbit as its spin angular momentum is negligible compared to the spin angular momentum of Salacia and the orbital angular momentum. 

The mass ratio $q$ ($q = m_s/M_p$) for Salacia-Actaea is unknown. Conserving the system mass ($M_{sys} = 486.1^{+7.6}_{-7.4} \times 10^{18}$ kg, Table \ref{tab:fits}), we allowed the density of Actaea to range from 0.5 g cm$^{-3}$ to the effective system density of $\rho = 1.38^{+0.22}_{-0.18}$ g cm$^{-3}$, corresponding to $0.013 \lesssim q \lesssim 0.037$; $\rho = 0.5$ g cm$^{-3}$ is representative of comet nuclei while the size-density relationship for TNOs (density increases with size) suggests Actaea is less dense than Salacia. In order for outward migration to occur, the initial rotation period of Salacia $T_{p0}$ must be shorter than the initial orbital period $T_{n0}$. For a test initial semimajor axis $a_0$, $T_{p0}$ is constrained by conservation of angular momentum. If $T_{p0}>T_{n0}$, outward migration is nonviable. In Figure \ref{figure:tidalevolution} (a), we show the $a_0-q$ parameter space where outward migration is viable by this simple argument. Our testing showed that the tidal locking time $t_{\text{lock}}$, the time it takes for Salacia to be spun down to the orbit period, is a strong function of $q$ and is virtually independent of $a_0$. \cite{Arakawa2021} found that impact-forming satellites usually circularize within the range $3R_p \leq a_0 \leq 8R_p$, so Figure \ref{figure:tidalevolution} (a) plots this range.

In Figure \ref{figure:tidalevolution} (b), we show the tidal locking time $t_{\text{lock}}$ as a function of the mass ratio $q$ and dissipation factor $Q$ (see Appendix \ref{tidalmethods}). We used $a_0$ = 9$R_p$ where all $q$ values allow outward migration. In the range $50 \leq Q \leq 1250$, $t_{\text{lock}} < 1$ Gyr for $0.013 \lesssim q \lesssim 0.037$. We note that $t_{\text{lock}}$ approaches the age of the solar system as $Q$ approaches 10000.

Further insight into the dynamical history of Salacia and Actaea can be understood by using the (136199) Eris-Dysnomia system as an analog. Previous work suggests that the Eris-Dysnomia system also exhibits doubly-synchronous rotation \citep{Szakats2023,Bernstein2023}. Working with this assumption, we show the same plots for the Eris-Dysnomia system in Figure \ref{figure:tidalevolution} (bottom row). Conserving the system mass ($M_{sys} = 16466 \pm 85 \times 10^{18}$ kg), we allowed the density of Dysnomia to range from 0.5 g cm$^{-3}$ to the effective system density of $\rho = 2.43 \pm 0.05$ g cm$^{-3}$, corresponding to $0.0055 \lesssim q \lesssim 0.027$ \citep{Holler2021}. Using the same range of dissipation factors $Q$, the tidal locking time $t_{\text{lock}}$ for Eris-Dysnomia is approximately an order of magnitude longer than that of Salacia-Actaea. In order for Eris-Dysnomia to synchronize within the the age of the solar system, this places an upper limit on $Q$ for a given $q$ (the tidal locking timescale $\tau \propto q^{-1}$, and $\tau \propto Q$). Allowing Eris and Dysnomia to have the same density ($q  \simeq 0.027$), $Q$ can be as large as $Q_{\text{Eris}} = 1175$. The 3$\sigma$ upper limit to the Dysnomia/Eris mass ratio is $q=0.015$ \citep{Brown2023}. From Figure \ref{figure:tidalevolution} (e), this places an upper limit on $Q$, such that $Q_{\text{Eris}} < 650$. For Salacia-Actaea, a $Q_{\text{Salacia}}=650$ corresponds to a tidal locking time-range of $150$ Myr $< t_{\text{lock}} < 400$ Myr. If Salacia's interior is as dissipative as that of Eris, Salacia's spin period became locked to the orbital period within $400$ Myr after Actaea was captured/formed. Even if Salacia is half as dissipative as Eris ($Q_{\text{Salacia}} = 1300$), $t_{\text{lock}} < 1.1$ Gyr. 

We note that the tidal model utilized in this work to examine rotation period evolution is simplistic; it assumes a homogeneous rock-ice composition and does not model the temporal evolution of the interior of the primary. \cite{Nimmo2023} evolved the Eris-Dysnomia system using a differentiated three-layer model of Eris consisting of a purely elastic 30 km-thick ice lithosphere on top of an iso-viscous 90 km-thick ice shell over a rigid silicate core. In that work, they set the tidal dissipation factor, $Q$, to be a function of the time-varying internal temperature of Eris as well as the evolving forcing period ($P_{\text{force}} = 2\pi / [\Omega (t) - n (t)]$, see Appendix \ref{tidalmethods}), where $Q$ is inversely proportional to temperature and $P_{\text{force}}$. Since Eris cooled down over time while the forcing period increased, they found that these two time-dependent effects on $Q$ roughly canceled each other out such that $Q$ was virtually constant with time at $Q \simeq 420$ when using a mass ratio $q = 0.0084$. For the constant-$Q$ model used in the present work, and using the same mass ratio, Figure \ref{tidalevolution} requires $Q_{\text{Eris}} \lesssim 400$ for the system to synchronize within the age of the solar system. We conclude that our simple tidal evolution model is adequate for estimating tidal locking timescales.


\section{Discussion}
\label{discussion}
Under the assumption that Salacia's dissipation factor, $Q_{\text{Salacia}}$, is approximately within an order of magnitude as that of Eris ($Q_{\text{Eris}} < 650$), our rudimentary tidal evolution model predicted that Salacia and Actaea would tidally lock in a doubly-synchronous state within the age of the solar system. If we let $Q_{\text{Salacia}} = Q_{\text{Eris}}$, the Salacia-Actaea tidal locking time $t_{\text{lock}}$ falls in the range $150$ Myr $< t_{\text{lock}} < 400$ Myr. Observations of the Salacia-Actaea system presented in this work support the hypothesis that they exhibit doubly-synchronous rotation. 

In light of these observations where long time-baseline photometry revealed a longer photometric period than much shorter baseline photometry, other TNO binary systems may be much more tidally evolved than short time-baseline photometry has indicated. We applied the same tidal evolution model and methodology to the (90482) Orcus-Vanth system, where recent lightcurve analysis of this TNO system suggested non-synchronous rotation \citep{Kiss2020}. Constraining the rotation period of Orcus by lightcurve analysis is challenging due to the small aspect angle at which we view the system. Assuming the spin pole of Orcus is within a few degrees of the mutual orbit pole, the average aspect angle over the past $\sim20$ yr has been $\psi \simeq 27 ^{\circ}$. This can be compared to the larger $\psi \simeq 51 ^{\circ}$ aspect angle at which we observe Salacia-Actaea. The mass ratio of the Orcus-Vanth system has been constrained by \cite{Brown2023}. Assuming the 3$\sigma$ lower limit to the mass ratio ($q = 0.12$), the spin periods of Orcus and Vanth synchronized within the age of the solar system even when we allowed the tidal dissipation factor to be an order of magnitude greater than the upper limit derived for Eris ($Q_{\text{Eris}} < 650$) (recall $\tau \propto q^{-1}$, and $\tau \propto Q$); Orcus and Vanth have likely tidally evolved to become fully synchronized as initially suggested in \cite{Brown2023}. 

We encourage future observations to test the hypothesis that the TNO systems Orcus-Vanth, (174567) Varda-Ilmare, as well as (38628) Huya and its satellite, are doubly synchronous. The (229762) G!k{\'u}n$||$'h{\`o}md{\'i}m{\`a}-G!{\`o}'{\'e} !H{\'u} system is of particular interest; the likely low mass fraction $q \simeq 0.011$ and large semimajor axis $a \simeq 19 R_p$ \citep{Grundy2019UK} suggests the synchronization timescale is approximately an order of magnitude longer than that of the Salacia-Actaea system, similar to the Eris-Dysnomia system. If this system is indeed synchronized, this would point to G!k{\'u}n$||$'h{\`o}md{\'i}m{\`a}'s interior being quite dissipative and would call for more complex thermal-orbit modeling \citep[e.g.][]{Nimmo2023, Akiba2025}.

Under the previously held inference that the Salacia-Actaea and Orcus-Vanth systems are not synchronized, as evidenced by past short time-baseline lightcurve analysis, \citet{Arakawa2025} showed that this was consistent with Orcus and Salacia having low rock mass fractions, similar to comets. Our finding that Salacia and Actaea are fully synchronized calls for future investigations into the rock mass fractions present in Salacia, Orcus, and other TNOs.

\newpage
\appendix

\section{Orbit Fitting}
\label{orbit1}

\begin{deluxetable*}{lcc}
\tabletypesize{\footnotesize}
\tablecaption{Keplerian Orbit Solution for Salacia-Actaea}
\tablehead{
Parameter &  & Posterior
}
\startdata
\textbf{\textit{Fitted parameters}}  &               &                                 \\
System mass ($10^{18}$ kg)           & $M_{sys}$     & $486.1^{+7.6}_{-7.4}$           \\
Semi-major axis (km)                 & $a$           & $5700^{+30}_{-29}$              \\
Eccentricity                         & $e$           & $0.008^{+0.003}_{-0.003}$       \\
Inclination ($\degr$)                & $i$           & $17.2^{+0.5}_{-0.5}$            \\
Argument of periapsis ($\degr$)           & $\omega$      & $41^{+33}_{-22}$                \\
Longitude of the ascending node ($\degr$)            & $\Omega$      & $108.9^{+1.6}_{-1.6}$           \\
Mean anomaly at epoch ($\degr$)               & $\mathcal{M}$ & $157^{+27}_{-22}$               \\
\hline
\textbf{\textit{Derived parameters}} &               &                                 \\
Orbit period (d)         & $P_{orb}$     & $5.49389^{+0.00001}_{-0.00001}$ \\
Orbit pole RA ($\degr$)              & $\alpha_{orb}$  & $313.7^{+1.2}_{-1.2}$                         \\
Orbit pole dec. ($\degr$)            & $\delta_{orb}$  & $66.1^{+0.5}_{-0.5}$                         \\
Orbit pole lon. ($\degr$)              & $\lambda_{orb}$  & $18.9^{+1.6}_{-1.6}$                         \\
Orbit pole lat. ($\degr$)            & $\beta_{orb}$  & $72.8^{+0.5}_{-0.5}$                         \\
\enddata
\tablecomments{Reported values represent the median value and uncertainties are based on 16th and 84th percentile values. All fitted angles are relative to the J2000 ecliptic plane on Salacia-centric JD 2454300 (2007 Jul 18 12:00 UT), except for RA and dec. values which are referenced to the J2000 equatorial coordinate system. }
\vspace{-1.15cm}
\label{tab:fits}
\end{deluxetable*}

Observations from Keck followed the same procedures as past observations \citep[e.g.,][]{Grundy2019}. As part of Keck program  N193 we used the laser guide star adaptive optics system \citep{wizinowich2006} with the NIRC2 narrow camera\footnote{\url{https://www2.keck.hawaii.edu/inst/nirc2}}. Images were taken with the $H$ filter, with wavelengths between $\sim$1.48 to 1.77 $\mu$m. Relative astrometry was extracted from flat-fielded pairs of subtracted dithered images with Gaussian PSF fitting, using the same techniques as a variety of studies in the TNO binary literature \citep{Stansberry2012,Grundy2019}. Observations from HST program 17848 were acquired with the Wide Field Camera 3 (WFC3) instrument using the F350LP filter. Calibrated images from MAST had astrometry extracted using PSF fitting with model WFC3 PSFs from TinyTim using well-validated techniques \citep[][]{Grundy2019}. We implement a 2-mas noise floor on the astrometry to account for any systematic errors in the astrometric reduction (i.e. time variable distortion of the NIRC2/WFC3 field, uncertainties in the distortion solution, uncertainties in pixel scale, etc.). 

Orbit fitting was accomplished using \texttt{MultiMoon}, an orbit fitting software designed to fit relative astrometry of TNO binaries \citep{ragozzine2024beyond}. \texttt{MultiMoon} is built on a Bayesian framework, approaching the orbit fitting problem as a task for Bayesian inference by using the \texttt{emcee} Markov Chain Monte Carlo (MCMC) sampler \citep{foreman2013emcee}. We use the Keplerian orbit fitting module to fit Salacia-Actaea's mutual orbit. Although recent studies have found that Actaea's orbit may be better fit by a precessing non-Keplerian orbit \citep{proudfoot2024beyond}, the approximation of a Keplerian orbit still provides an extremely sensitive way to measure the mutual orbit period, system mass, and orbit orientation. We revisit the putative non-Keplerian nature of Actaea's orbit later. 

Orbit fits were run using standard \texttt{MultiMoon} fitting procedures. We used an ensemble of 960 walkers which were run for 15000 burn-in steps split between 10000 initial steps and 5000 post-pruning burn-in steps. The posterior distribution was then sampled using 10000 steps, producing converged fits. For more specific details on \texttt{MultiMoon} procedures and methods, see \cite{ragozzine2024beyond} and \cite{proudfoot2024beyond}.

\begin{figure}
\centering
 	\includegraphics[width=0.48\textwidth]{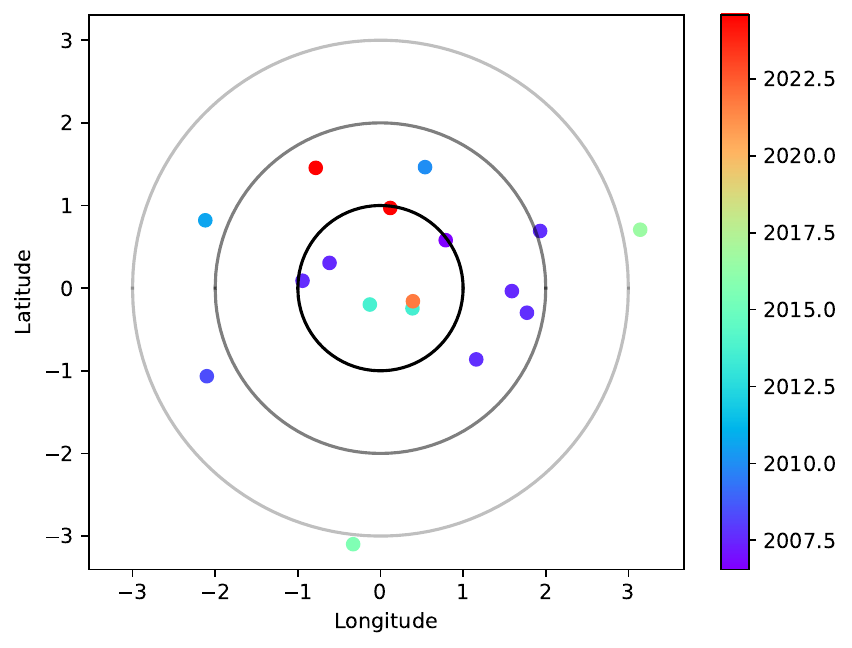}
	\caption{Two-dimensional view of the orbit fitting residuals. The colorbar indicates the date of observation.}
    \label{fig:resid}
\end{figure}

The results of our Keplerian orbit fits are shown in Table \ref{tab:fits}. Our fits achieve a best-fit $\chi^2=51.8$ with 26 degrees of freedom, giving a $\chi^2$ per degree of freedom of $2.0$, indicating a somewhat poor fit quality. We calculate the probability that random chance could give an equivalent (or worse) fit quality as 0.2\%. A two-dimensional view of the orbit fitting residuals is shown in Figure \ref{fig:resid}. However, even with such a poor fit quality, typical residuals in astrometry are of order $\sim$10 mas, showing that the Keplerian orbit provides an excellent approximation of Actaea's orbital motion. In our fits, we provide a new estimate of Actaea's orbit period, with uncertainties of $\sim$1 second. 

Our fits show that Actaea's prograde orbit seems to be eccentric, detecting a non-zero eccentricity at $\sim$2.5$\sigma$ confidence. Given the system's tidally evolved state, this small undamped eccentricity is quite surprising, with estimates of the eccentricity damping timescale of $<$20 Myr \citep{Stansberry2012}. Two plausible scenarios could explain this: first, a recent perturbation---from an impact or a TNO flyby---could have excited the mutual orbit's eccentricity, which has yet to be fully damped. Alternatively, small time-variable offsets between Salacia's (or Actaea's) center-of-body and center-of-light (COB-COL) could masquerade as a small orbital eccentricity. A similar residual eccentricity plagued Pluto-Charon orbit fits for decades, until accurate albedo maps of Pluto and Charon's surfaces were convolved with precise HST astrometry \citep{tholen1997orbit,buie2012orbit}. Given the system's single-peaked light curve, which appears to be the result of rotating albedo features (see below), this explanation seems a plausible---and importantly, parsimonious---explanation for the residual eccentricity detected in our orbit fits. 

With our measured eccentricity, we can estimate the size of these COB-COL offsets, assuming that these offsets are responsible for the entire eccentricity. The radial difference between apoapse and periapse is given as $\Delta r = 2ae \approx 91$ km. The required COB-COL offset would be a periodic function (phased to the mutual orbital motion) with an amplitude of $\Delta r/2 \approx 45$ km, or $\sim$5\% of Salacia's diameter. In comparison, Pluto's average COB-COL offset is $\sim$10\% its diameter \citep{buie2010pluto}. 

COB-COL variations may also be responsible for the poor fit quality found with a Keplerian orbit fit. Past works have suggested that the poor fit quality was due to non-Keplerian precession \citep{proudfoot2024beyond}. Importantly, however, in a tidally evolved system, non-Keplerian effects are generally minimized. Since the mutual orbital motion of Salacia and Actaea contain the majority of the system's angular momentum, misalignments between Salacia's rotation pole and the mutual orbit pole will tend to induce precession/libration of Salacia's pole, effectively eliminating nodal precession. Since eccentricity is quickly damped (as discussed above), apsidal precession is also minimized. In contrast, COB-COL offsets, which move in phase with the system's mutual orbital motion can appear as a small, precessing eccentricity.

\section{Photometry Calibration}
\label{photometry1}
We utilized the Massive prOcessing Of aStronomical imagEs (Moose) - version 2 \citep[M2;][]{M2} software to process FITS images from each observing run and output photometry of the Salacia-Actaea system. The output photometry is in the Johnson-Cousins $R$-band calibrated to the \cite{Landolt1992} photometric system.

Photometric calibrations in M2 are done on an image-by-image basis as follows. M2 makes use of in-frame stars with known apparent magnitudes in the Gaia Early Data Release 3 catalog, along with the Gaia $\to$ Johnson $V$ and Johnson $R$-band filter transformation equations published in \cite{Riello2021Gaia}. The majority of our observations were done through no filter at all, so transformation equations had to be computed between the filter/camera response and the \cite{Landolt1992} photometric system. Filter transformation equations are a function of some stellar (and TNO) color index, and since most large TNOs have precise $V-R$ colors, M2 uses filter transformation equations that are measured using the $V-R$ color for main-sequence type stars. 

For each in-frame star within the Gaia catalog, the $V$- and $R$-band magnitudes are computed using the known filter transformation equations from \cite{Riello2021Gaia}, along with the $V-R$ color. We refer to the $R$-band catalog magnitudes as $R_{cat}$. Instrumental fluxes are then measured for each star using basic aperture photometry, with source aperture and background-annulus radii related to the measured stellar FWHM (1.3 $\times$ FWHM radius for the source-aperture). For an instrumental magnitude of $m_{in} = - 2.5\text{log}(F)$, the instrumental magnitude and $R_{cat}$ are related by $R_{cat} - m_{in} = -Z + f(V-R)$, where $Z$ is the zeropoint and $f(V-R)$ is a third-order polynomial as a function of $V-R$ with coefficients $B_1$, $B_2$ and $B_3$. The polynomial coefficients and zeropoint are fitted with in-frame stars. For the TNO with flux $F_{TNO}$ and known color $(V-R)_{TNO}$ its $R$-band magnitude on the \cite{Landolt1992} system is

\begin{equation}
\begin{split}
R_{\text{TNO}} =  2.5\text{log}(F) - Z \hspace{3.5cm}\\
+ B_1(V-R) + B_2(V-R)^2 + B_3(V-R)^3  \Big{|}_{\text{TNO}}. 
\end{split}
\end{equation}

M2 has various stages that filter out poor images and poor calibration stars, and ensures minimal contamination in the TNO flux. After M2 tabulated the photometry, we used additional filters and checks to ensure accurate photometry of the Salacia-Actaea system. We required there to be $\geq 10$ calibration stars in each image. The color-coefficients in the filter transformation equations for a given filter/camera combination should be constant in nature from night-to-night, image-to-image. As such, we tabulated photometry for each filter/camera combination and ran a sigma-clipping routine for each of the color coefficients (at 3$\sigma$), filtering out observations where the color-coefficients were extremal. Lastly, we ensured the apertures used for the photometry were centered on the expected location of the TNO, and that it tracked the TNO's expected position over the course of each observing run and did not jump to nearby sources.


\section{Tidal Evolution Model}
\label{tidalmethods}
Assuming Salacia's spin pole is parallel to the orbit pole and that angular momentum is conserved, Salacia's rotation period and the primaricentric semi-major axis evolve according to \citep{Goldreich1966,Murray2000}

\begin{equation}
\label{eq:spin}
    \frac{d \Omega}{dt} = \text{sign}(n - \Omega) \frac{45 G}{8}\left(\frac{R_p^3 m_s^2}{M_p} \frac{1}{Q'}\right) \left(\frac{1}{a^6}\right)
\end{equation}

\begin{equation}
\label{eq:orbit}
    \frac{da}{dt}       =-\text{sign}(n - \Omega) \frac{9\sqrt{G}}{2} \left(\frac{R_p^5 m}{\sqrt{M}} \frac{1}{Q' } \right) \left(\frac{1}{a^{11/2}}\right),
\end{equation}

\noindent where $\Omega$ is Salacia's spin angular velocity $n$ is the mean motion, $G$ is the gravitational constant, $R_p$ is the radius of Salacia, $M_p$ is the mass of Salacia, $m_s$ is the mass of Actaea, $a$ is the primaricentric semi-major axis, and 

\begin{equation}
    Q' = Q\left(1 + \frac{57}{8\pi G} \frac{\mu}{R_p^2 \rho_p^2}\right)
\end{equation}

\noindent for dimensionless tidal dissipation factor $Q$, modulus of rigidity $\mu$ and mass-density $\rho$. Salacia and Actaea's diameters are $D_p = 866 \pm 37$ km and $D_s = 290 \pm 21$ km, respectively \citep{Brown2017}. The total mass is $M_{sys} = (486.1 \pm 7.5) \times 10^{18}$ kg (Table \ref{tab:fits}). 

For the modulus of rigidity, we used $\mu=4$ GPa which is representative of water ice \citep{Neumeier2018}. Although a rigidity of $\mu \simeq 67$ GPa (typical of solid rock) has been used in the past, Salacia's bulk material properties during the initial $\sim$1000 Myr when tidal evolution occurred is not well modeled as a solid monolith. We note that if Salacia differentiated sufficiently fast (within $\sim$100 Myr) and hosts a large monolithic rocky core, this would decrease the dissipation in the interior, increasing the tidal evolution timescales. Indeed, Salacia today may be differentiated with a rocky core, however this required an epoch of melting, where water percolated through the pores of a rocky matrix before refreezing \citep{Loveless2022}.

In this work, the evolution of the Salacia-Actaea system was evaluated numerically using Equations \ref{eq:spin} \& \ref{eq:orbit} and Euler's method. This numerical method has several benefits, including (a) allowing multiple changes in the sign($n-\Omega$) argument during evolution and (b) conservation of the physical parameters that control the migration timescale.

\newpage
\begin{deluxetable*}{cccCCCC}
\tablewidth{\textwidth}
\tablecaption{Observed Astrometric Positions of Actaea}
\tablehead{
Julian Date & Date & Telescope/Camera & \Delta x & \Delta y & \sigma_{\Delta x} & \sigma_{\Delta y} \\
 & & & ('') & ('') & ('') & ('')
}
\startdata
2453938.41493 & 2006/07/21 & HST/ACS-HRC  & +0.04706  & -0.10224 & 0.00200 & 0.00200 \\
2454295.37924 & 2007/07/13 & HST/WFPC2-PC & +0.02193  & -0.11084 & 0.00391 & 0.01344 \\
2454297.32976 & 2007/07/15 & HST/WFPC2-PC & +0.12371  & +0.09208  & 0.00448 & 0.00206 \\
2454323.96413 & 2007/08/11 & HST/WFPC2-PC & +0.17781  & +0.00363  & 0.00200 & 0.00303 \\
2454324.77860 & 2007/08/12 & HST/WFPC2-PC & +0.12785  & +0.09402  & 0.00391 & 0.00200 \\
2454324.90898 & 2007/08/12 & HST/WFPC2-PC & +0.10084  & +0.09889  & 0.00863 & 0.00316 \\
2454346.90493 & 2007/09/03 & HST/WFPC2-PC & +0.10314  & +0.09949  & 0.00358 & 0.00347 \\
2454606.27369 & 2008/05/19 & HST/WFPC2-PC & -0.10096 & +0.07422  & 0.00200 & 0.00200 \\
2455176.77133 & 2009/12/11 & Keck 2/NIRC2 & +0.05801  & +0.10558  & 0.00613 & 0.00300 \\
2455411.90558 & 2010/08/03 & Keck 2/NIRC2 & +0.18274  & +0.01839  & 0.00300 & 0.00300 \\
2456492.02668 & 2013/07/18 & Gemini/NIRI  & -0.19689 & -0.07799 & 0.10000 & 0.10000 \\
2456529.01220 & 2013/08/24 & Gemini/NIRI  & -0.06779 & +0.11576  & 0.10000 & 0.10000 \\
2457234.99390 & 2015/07/31 & Keck 2/NIRC2 & +0.05868  & -0.08215 & 0.00312 & 0.00654 \\
2457606.96819 & 2016/08/06 & Keck 2/NIRC2 & -0.18064 & -0.05189 & 0.00300 & 0.00300 \\
2459453.88191 & 2021/08/27 & Keck/NIRC2   & -0.07667 & -0.12512 & 0.00223 & 0.00200 \\
2460521.40133 & 2024/07/29 & HST/WFC3     & +0.16572  & +0.01519  & 0.00200 & 0.00200 \\
2460524.23293 & 2024/08/01 & HST/WFC3     & -0.16554 & -0.03485 & 0.00200 & 0.00224 \\
\enddata
\tablecomments{$x$ and $y$ correspond to Right Ascension and Declination, respectively. Uncertainties on $\Delta x,y$ have noise floors of 2 mas.}
\label{tab:observations}
\end{deluxetable*}

\begin{acknowledgments}

This research is based on observations made with the NASA/ESA Hubble Space Telescope obtained from the Space Telescope Science Institute, which is operated by the Association of Universities for Research in Astronomy, Inc., under NASA contract NAS 5–26555. These observations are associated with program 17848.

Some of the data presented herein were obtained at Keck Observatory, which is a private 501(c)3 non-profit organization operated as a scientific partnership among the California Institute of Technology, the University of California, and the National Aeronautics and Space Administration. The Observatory was made possible by the generous financial support of the W. M. Keck Foundation. 

This work has made use of data from the European Space Agency (ESA) mission {\it Gaia} (\url{https://www.cosmos.esa.int/gaia}), processed by the {\it Gaia} Data Processing and Analysis Consortium (DPAC,
\url{https://www.cosmos.esa.int/web/gaia/dpac/consortium}). Funding for the DPAC has been provided by national institutions, in particular the institutions participating in the {\it Gaia} Multilateral Agreement.

The authors wish to recognize and acknowledge the very significant cultural role and reverence that the summit of Maunakea has always had within the Native Hawaiian community. We are most fortunate to have the opportunity to conduct observations from this mountain. 

C.C. and E.F.V. acknowledge the support of NASA Solar System Workings grant 19-SSW19-0190. C.C. acknowledges support from the Florida Space Grant Consortium. B.P. and F.L.R. acknowledge the support of the University of Central Florida Preeminent Postdoctoral Program (P$^3$). 
\end{acknowledgments}

\clearpage
\bibliography{zreferences}{}
\bibliographystyle{aasjournalv7}

\end{document}